\documentclass[5p, times, singlecolumn]{article}
\usepackage[T1]{fontenc}
\usepackage{doi}
\usepackage{units}

\usepackage{chemformula}
\setchemformula{kroeger-vink}

\begin{document}
\begin{center}

\LARGE Influence of Br$^{-}$/S$^{2-}$ site-exchange on Li diffusion mechanism in Li$_6$PS$_5$Br - a computational study

\bigskip

\normalsize
Marcel Sadowski and Karsten Albe\\
Email: sadowski@mm.tu-darmstadt.de and albe@mm.tu-darmstadt.de

\bigskip
\
Institute of Materials Science, Technical University of Darmstadt, 64287 Darmstadt, Germany

\end{center}

\begin{abstract}
We investigate the influence of  Br$^-$/S$^{2-}$ site-exchange on lithium diffusion in the agyrodite-type solid electrolyte Li$_6$PS$_5$Br by ab-initio molecular dynamics simulations.
Based on the calculated trajectories a new mechanism for the internal lithium reorganization within the Li-cages around the $4d$ sites is identified.
This reorganization mechanism is highly concerted and cannot be described by one single rotation axis only.
Simulations with Br$^-$/S$^{2-}$ defects reveal that \ch{Li_i^.} interstitials are the dominant mobile charge carriers, which
originate from Frenkel pairs. 
These are formed because  \ch{Br_S^.} defects on the $4d$ sites cause the transfer of one or even two \ch{Li_i^.}  to the neighboring 12 cages.
The lithium interstitials then carry out intercage jumps via interstitial and interstitialcy mechanisms.
With that, one single \ch{Br_S^.} defect enables Li diffusion over an extended spatial area explaining why low degrees of site-exchange are sufficient to trigger superionic conduction.
The vacant sites of the Frenkel pairs,  namely  \ch{V^'_{Li}},  are mostly immobile and bound to the \ch{Br_S^.} defect.
To a lesser degree also \ch{S_{Br}'} defects induce disturbances in the lithium distribution and act as sinks for lithium interstitials restricting the  \ch{Li_i^.} motion to the vicinity of the \ch{S_{Br}'} defect.
\end{abstract}

\section{Introduction}

\begin{figure}[ht!]
    \centering
    \includegraphics[width=0.95\textwidth]{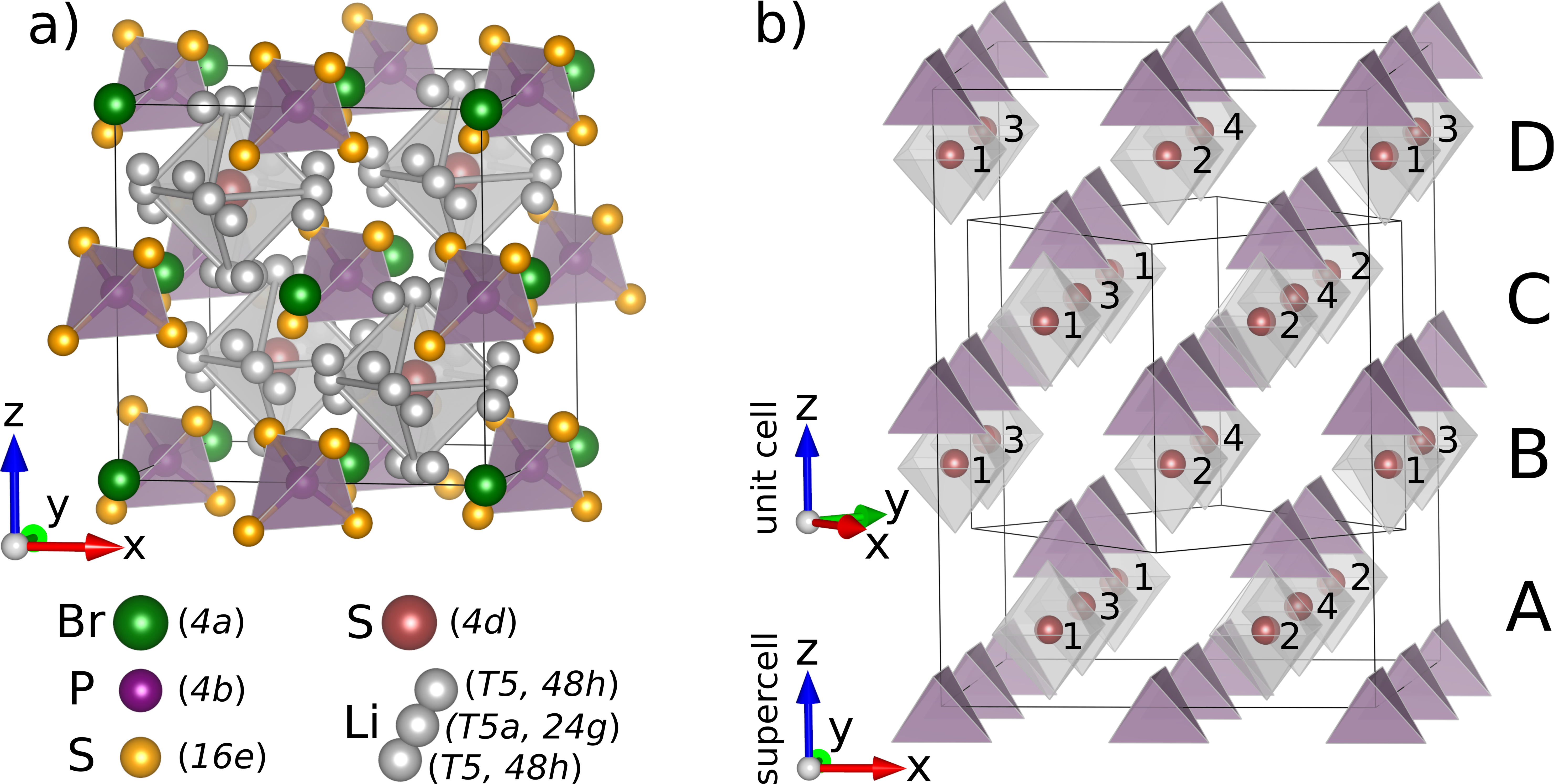}
    \caption{a) Basic unit cell of Li$_6$PS$_5$Br. The octahedral Li cages that are spanned by the Li T5a sites around $4d$ are indicated. b) Used $\sqrt{2}\times\!\!\sqrt{2}\times2$ supercell model for all AIMD simulations. For simplicity only PS$_4^{3-}$ tetrahedra and Li octahedra around the $4d$ sites are shown. In the center of the supercell one original unit cell is indicated. Unique names A1, A2, ... , D4 have been assigned to each nominal $4d$ sites as labeled in the figure. Their fractional coordinates can be read from Table~\ref{tab:Cage_center_names}.}
    \label{fig:Figure_1_Li6PS5Br_structure_setup}
\end{figure}

Sulfide solid electrolytes (SE) are promising candidates to be used in lithium all-solid-state batteries due to their high ionic conductivity that competes with conventional liquid organic electrolytes.
Especially, argyrodite-type Li$_6$PS$_5$X (X = Cl, Br, I) materials first proposed by Deiseroth et al. in 2008~\cite{deiseroth_li6ps5x_2008} have recently attracted much attention~\cite{deiseroth_li7ps6_2011,kraft_influence_2017,yu_accessing_2017,hanghofer_fast_2019}.
Their basic crystal structure is cubic with $F\bar{4}3m$ symmetry (Figure~\ref{fig:Figure_1_Li6PS5Br_structure_setup}a).
The rock-salt arrangement of anions can be described as a face-centered cubic lattice of X$^-$ anions (Wyckoff $4a$) and a shifted face-centered cubic lattice of PS$_4^{3-}$ units (P ion centered on Wyckoff $4b$).
Half of the tetrahedral sites are occupied by S$^{2-}$ (Wyckoff $4d$).
We note that the nomenclature of the $4d$ sites varies between different reports and is sometimes denoted as $4c$, which originates from different choices of the origin of the coordinate system.
The structure accommodates various Li$^+$ sites with partial occupancies.
Forming an octahedron, each X$^-$ on $4d$ is surrounded by six Li$^+$ sites (Wyckoff $24g$, also called T5a).
Additionally, all T5a sites are surrounded by two Li$^+$ sites (Wyckoff $48h$, also called T5).
Due to the close distances inside such a T5-T5a-T5 triplets it is often argued, that it is only occupied by one Li$^+$ at a time.
Between the triplets another site (Wyckoff $48$, also called T2) was recently reported~\cite{minafra_local_2020}.

Based on this structure three main jump types have been identified: (1) motion inside a T5-T5a-T5 triplet, also referred to as doublet jump, (2) intracage jumps between two different T5-T5a-T5 triplets of the same cage and (3) intercage jumps between T5-T5a-T5 triplets of neighboring cages.
The doublet jump has a small migration barrier and the T5-T5a-T5 triplets can also be viewed as a local basin of sites with a flat potential energy surface.
The intracage jump was found to be highly correlated and two mechanism, a rigid octahedral rotation and a trigonal-prismatic internal reorganization, have been identified recently~\cite{morgan_chem_2020}.
Effective Li transport, however, can only be achieved via intercage jumps. 
These have been reported to be the bottleneck for long-range diffusion~\cite{yu_unravelling_2016,yu_accessing_2017} and seem to be facilitated by the T2 sites~\cite{minafra_local_2020}.

The basic crystal structure of Li$_6$PS$_5$X can be modified in different ways in order to enhance the materials properties. 
Several approaches such as (i) varying  the X$^-$/S$^{2-}$ content towards Li$_{6-n}$PS$_{5-n}$X$_{1+n}$ (influence on Li content), (ii) replacing S with elements such as Se or Te (influence on the lattice softness and polarizability), (iii) replacing P with aliovalent elements such Si, Ge or Sn (influence on Li content) and/or (iv) replacing Li with Na have been successfully applied~\cite{chen_stability_2015,wang_theoretical_2017,zhu_li3yps42_2017,adeli_boosting_2019,ouyang_computational_2020,wang_fast_2020,zhang_silicon-doped_2020}.
However, even in the non-modified Li$_6$PS$_5$X systems (X = Cl, Br) the basis structure can be influenced and structural site-disorder between X$^-$/S$^{2-}$ on the $4a$/$4d$ sites is achieved depending on the synthesis conditions.
For Li$_6$PS$_5$Br this was recently shown using
quenching experiments from elevated temperatures. 
The quenching kinetically freezes a certain degree of X$^-$/S$^{2-}$ site-exchange that is present at higher temperatures without the need of changing the composition~\cite{gautam_rapid_2019}.
The site-exchange influences the Li substructure which in turn leads to enhanced transport properties.

In summary, a variety of optimization strategies have already been successfully applied to Li$_6$PS$_5$X and related systems.
In this regard, atomistic computer simulations helped to understand the material and the influence of the optimization strategies~\cite{pecher_atomistic_2010,rao_studies_2011,rayavarapu_variation_2012,holzwarth_computer_2011,de_klerk_diffusion_2016,yu_unravelling_2016,wang_theoretical_2017,deng_data-driven_2017,stamminger_ionic_2019,gautam_rapid_2019,ouyang_computational_2020,morgan_chem_2020}.
For the development of further reliable optimization strategies, however, a even more detailed understanding of the atomistic processes is necessary.

In this contribution we are using ab-initio molecular dynamics simulations (AIMD) to study the Li diffusion mechanism and how it is influenced by the X$^-$/S$^{2-}$ site-exchange on the $4a$/$4d$ sites.
We restrict the analysis to Li$_6$PS$_5$Br, but the generated knowledge should be transferable to other halide ions and will also be helpful for the development and optimization of other superionic conductors whose properties are largely influenced by structural disorder.

\section{Computational Details}
Density Functional Theory (DFT) calculations have been performed using Version 5.4.4.~of the \textit{Vienna ab-initio simulation package} (VASP)~\cite{vasp1,vasp2,vasp3,vasp4} with plane-wave basis sets and the Perdew-Burke-Ernzerhof (PBE) exchange-correlation functional~\cite{pbe1,pbe2}.
Projector-augmented-wave (PAW)~\cite{paw1,paw2} pseudopotentials for Li, P, S and Br have been used as provided by VASP that explicitly treat the outer 3, 5, 6 and 7 electrons as valence electrons, respectively.
The pseudopotential versions were PAW\_PBE Li\_sv 10Sep2004, PAW\_PBE P 06Sep2000, PAW\_PBE S 06Sep2000 and PAW\_PBE Br 06Sep2000.

During geometry optimizations of atomic structures the cutoff energy for the plane-wave basis set was 600~eV. 
The convergence criteria for the energy of self-consistent electronic calculations was 10$^{-6}$~eV and 0.01~eV/\AA\ for atomic forces.
The k-spacing was set to 0.25~\AA$^{-1}$ resulting in a $3\times3\times3$ k-mesh for one unit cell as shown in Figure~\ref{fig:Figure_1_Li6PS5Br_structure_setup}a.
In order to get an initial starting structure only the T5a Li sites on Wyckoff $24g$ were occupied in a unit cell of Li$_6$PS$_5$Br without Br$^-$/S$^{2-}$ site-exchange.
The atomic positions and the simulation box volume were then optimized while the box was kept cubic. 
The resulting lattice constants of 10.2855~\AA\ was used as basis for all cells described in the following.
Next, a $\sqrt{2}\times\!\!\sqrt{2}\times2$ supercell containing 96 Li ions was generated and the \textit{Supercell Program}~\cite{okhotnikov_supercell_2016} was used to create structures with different degrees of Br$^-$/S$^{2-}$ site-exchange.

These structures were used as starting structures for ab-initio molecular dynamics (AIMD) simulations using VASP.
For AIMD calculations the cutoff energy for the plane-wave basis set was reduced to the default value as specified by the pseudopotentials and the convergence criterion for the self-consistent electronic calculations was set to 10$^{-5}$~eV.
Calculations were performed only at the gamma point with timesteps of 1~fs.

Additionally, a S-rich system containing a \ch{S'_{Br}} defect was generated by placing a S on a Br $4a$ site in the supercell model. 
One interstitial Li (\ch{Li_i^.}) was added in the vicinity of the substituted ion to ensure charge neutrality.
Similarly, also a Br-rich system with a \ch{Br_S^.} defect was generated that was compensated by a Li vacancy \ch{V_{Li}'}.

By inspection of the Li trajectories event lists of the jump processes have been extracted and are presented as flow charts.
Atomic structures are visualized using \textit{VESTA}~\cite{momma_vesta_2011} and \textit{Ovito}~\cite{stukowski_visualization_2010}.

\begin{table}[ht!]
    \centering
    \caption{Fractional coordinates $x,y,z$ of the labeled $4d$ sites A1, A2, ... , D4 as shown in Figure~\ref{fig:Figure_1_Li6PS5Br_structure_setup}b.}
    \begin{tabular}{c||c|c|c|c}
         & 1 & 2 & 3 & 4  \\
         \hline 
         \hline
      A & $\nicefrac{1}{4},0,\nicefrac{1}{8}$   &  $\nicefrac{3}{4},0,\nicefrac{1}{8}$ & $\nicefrac{1}{4},\nicefrac{1}{2},\nicefrac{1}{8}$  &  $\nicefrac{3}{4},\nicefrac{1}{2},\nicefrac{1}{8}$ \\\hline
      B & $0,\nicefrac{1}{4},\nicefrac{3}{8}$   & $\nicefrac{1}{2},\nicefrac{1}{4},\nicefrac{3}{8}$  & $0,\nicefrac{3}{4},\nicefrac{3}{8}$  &  $\nicefrac{1}{2},\nicefrac{3}{4},\nicefrac{3}{8}$ \\\hline
      C  & $\nicefrac{1}{4},0,\nicefrac{5}{8}$   &  $\nicefrac{3}{4},0,\nicefrac{5}{8}$ & $\nicefrac{1}{4},\nicefrac{1}{2},\nicefrac{5}{8}$  &  $\nicefrac{3}{4},\nicefrac{1}{2},\nicefrac{5}{8}$ \\\hline
      D  & $0,\nicefrac{1}{4},\nicefrac{7}{8}$   & $\nicefrac{1}{2},\nicefrac{1}{4},\nicefrac{7}{8}$  & $0,\nicefrac{3}{4},\nicefrac{7}{8}$  &  $\nicefrac{1}{2},\nicefrac{3}{4},\nicefrac{7}{8}$ \\
      \end{tabular}
    \label{tab:Cage_center_names}
\end{table}

\section{Results and Discussion}

\subsection{Li diffusion without site-exchange}
In order to discuss how Br$^-$/S$^{2-}$ site-exchange on $4a$/$4d$ sites affects Li$^+$ diffusion the non-disturbed system without site-exchange needs to be analyzed first as a reference.
Thus, we conducted ab-initio molecular dynamics (AIMD) simulations using a $\sqrt{2}\times\!\!\sqrt{2}\times2$ supercell (see Figure~\ref{fig:Figure_1_Li6PS5Br_structure_setup}b) of Li$_6$PS$_5$Br without Br$^-$/S$^{2-}$ site-exchange. 
Two independent runs of approximately 60~ps length have been performed for improved statistics.
Unless specified differently, all  results presented in the following refer to simulations conducted at 600~K.

\begin{figure}[ht!]
    \centering
    \includegraphics[width=0.95\textwidth]{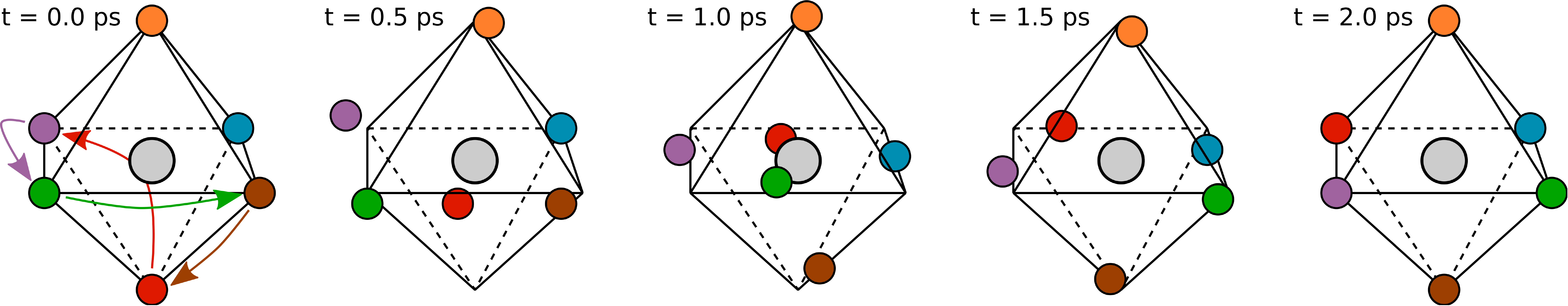}
    \caption{Schematic snapshots of the  newly observed internal reorganization mechanism within the Li$^+$ cages around the $4d$ sites. Arrows indicate the different diffusion paths. For clarification, Li$^+$ are illustrated as differently colored spheres. The large gray sphere in the center of the octahedron represents the $4d$ site. For better comparison, the form of representation was adopted from reference~\cite{morgan_chem_2020}.}
    \label{fig:ReorganisationMech}
\end{figure}

The Li$^+$ motion is dominated by doublet and intracage jumps.
The observed intracage jumps are based on two recently reported mechanisms, the rigid octahedral rotation and trigonal-prismatic reorganization mechanisms~\cite{morgan_chem_2020}.
Besides these two types of reorganization mechanisms we furthermore identified a third type which is depicted in Figure~\ref{fig:ReorganisationMech}.
This mechanism cannot be described by only one rotation axis and involves four Li$^+$.
During the reorganization, however, also the remaining Li$^+$ partly show extended displacements (see blue sphere in Figure~\ref{fig:ReorganisationMech}).
The whole process takes place within approximately 2.0~ps at 600~K and substantiates that Li$^+$ motion inside the cages is highly concerted.

\begin{figure}[ht!]
    \centering
    \includegraphics[width=0.99\textwidth]{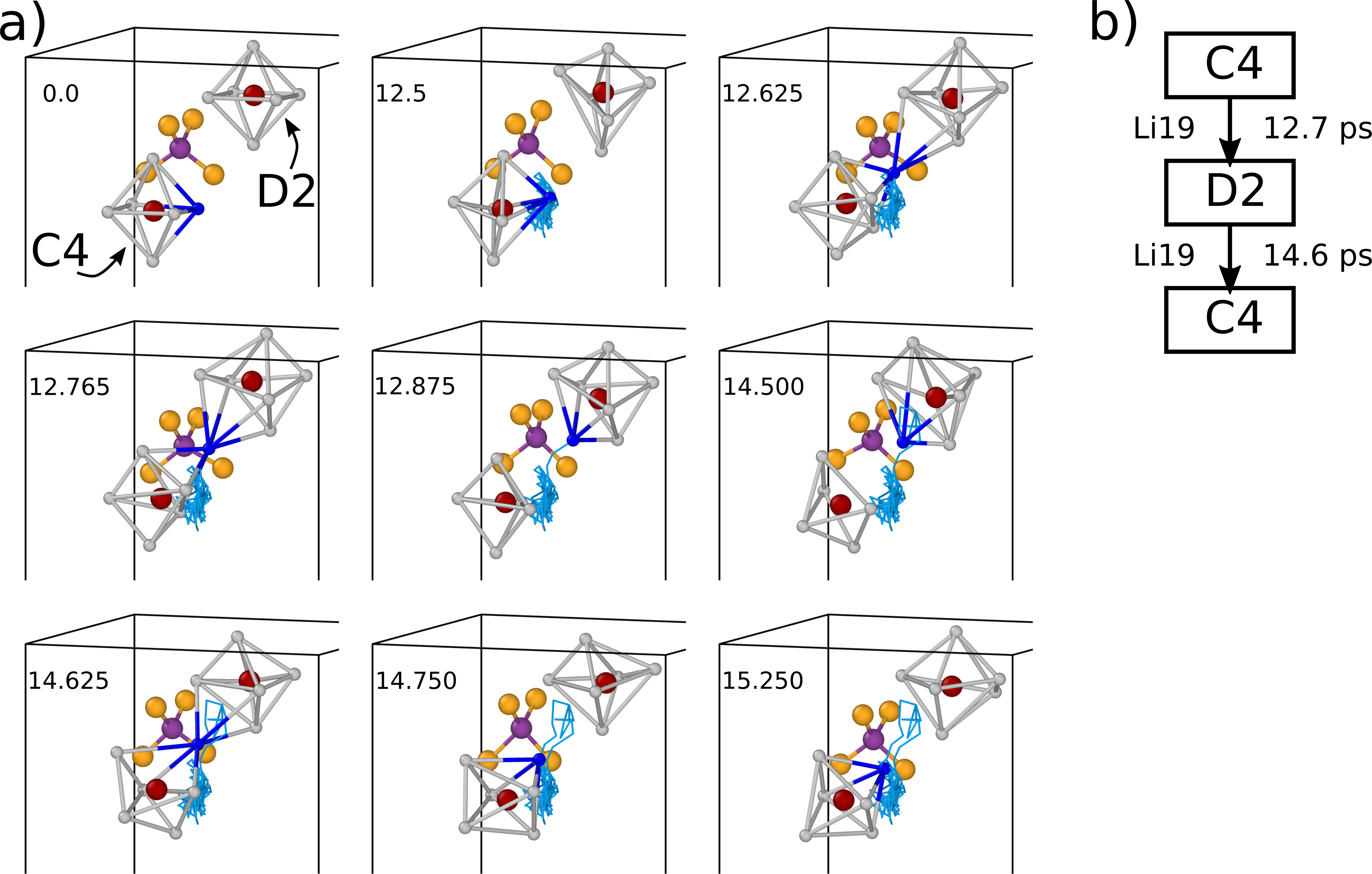}
    \caption{a) Snapshots of the intercage jump observed in the supercell without Br$^-$/S$^{2-}$ site-exchange. The jumping Li$^+$ is shown as deep blue sphere together with its trajectory in light blue and the simulation time in ps is labeled in the individual snapshots. The remaining color coding is the same as in Figure~\ref{fig:Figure_1_Li6PS5Br_structure_setup}a and the majority of ions is not shown for clear visibility. b) Flow chart notation of the shown jump event. The labels C4 and D2 refer to Li$^+$ cages around the corresponding $4d$ sites as defined in Figure~\ref{fig:Figure_1_Li6PS5Br_structure_setup}b.}
    \label{fig:JumpWithoutDisorder}
\end{figure}

In contrast to the high mobility of Li$^+$ within the cages the observation of intercage jumps is rare. 
During the two simulations only one single intercage jump event was observed which is shown in Figure~\ref{fig:JumpWithoutDisorder}a. 
At a simulation time of approximately 12.7~ps one Li$^+$ jumps from the cage centered around position C4 to the cage around D2 creating a situation in which the number of Li$^+$ around C4 and D2 are 5 and 7, respectively.
As normally the number of Li$^+$ would be 6 in both cages this configuration can be seen as a Frenkel pair defect with a Li$^+$ vacancy (in Kr\"oger-Vink notation: \ch{V_{Li}'}) around C4 and a Li$^+$ interstitial (\ch{Li_i^.}) around D2.
The unfavorable Li$^+$ distribution is restored by a back jump of the same Li$^+$ only 2~ps later demonstrating the highly correlated character of intercage jumps in the absence of Br$^-$/S$^{2-}$ site-exchange.
Only at temperatures above 700~K a sufficient number of intercage jumps within reasonable simulation time scales can be observed~\cite{stamminger_ionic_2019,gautam_rapid_2019}.

The situation changes as soon as Br$^-$/S$^{2-}$ site-exchange is introduced into the system.
In order to analyze the effect of Br$^-$/S$^{2-}$ site-exchange on the underlying atomistic processes we will in the following first divide the site-exchange into its individual parts: a substitutional S$^{2-}$ on nominal Br$^-$ $4a$ sites (\ch{S_{Br}'}) and a substitutional Br$^{-}$ on nominal S$^{2-}$ $4d$ sites (\ch{Br_{S}^.}) where charge compensation was realized by adding and removing Li, respectively.

Furthermore, Figure~\ref{fig:JumpWithoutDisorder}b introduces a flow chart notation that condenses the jump event described above in a short and comprehensible manner.
It contains the initial and final positions of the jumping Li$^+$, its index (here: Li19) and the approximate simulation time of the event.
The flow chart notation will later be used to depict the jump events observed in other simulations.

\subsection{Li diffusion induced by S on $4a$}

\begin{figure}[ht!]
    \centering
    \includegraphics[width=0.75\textwidth]{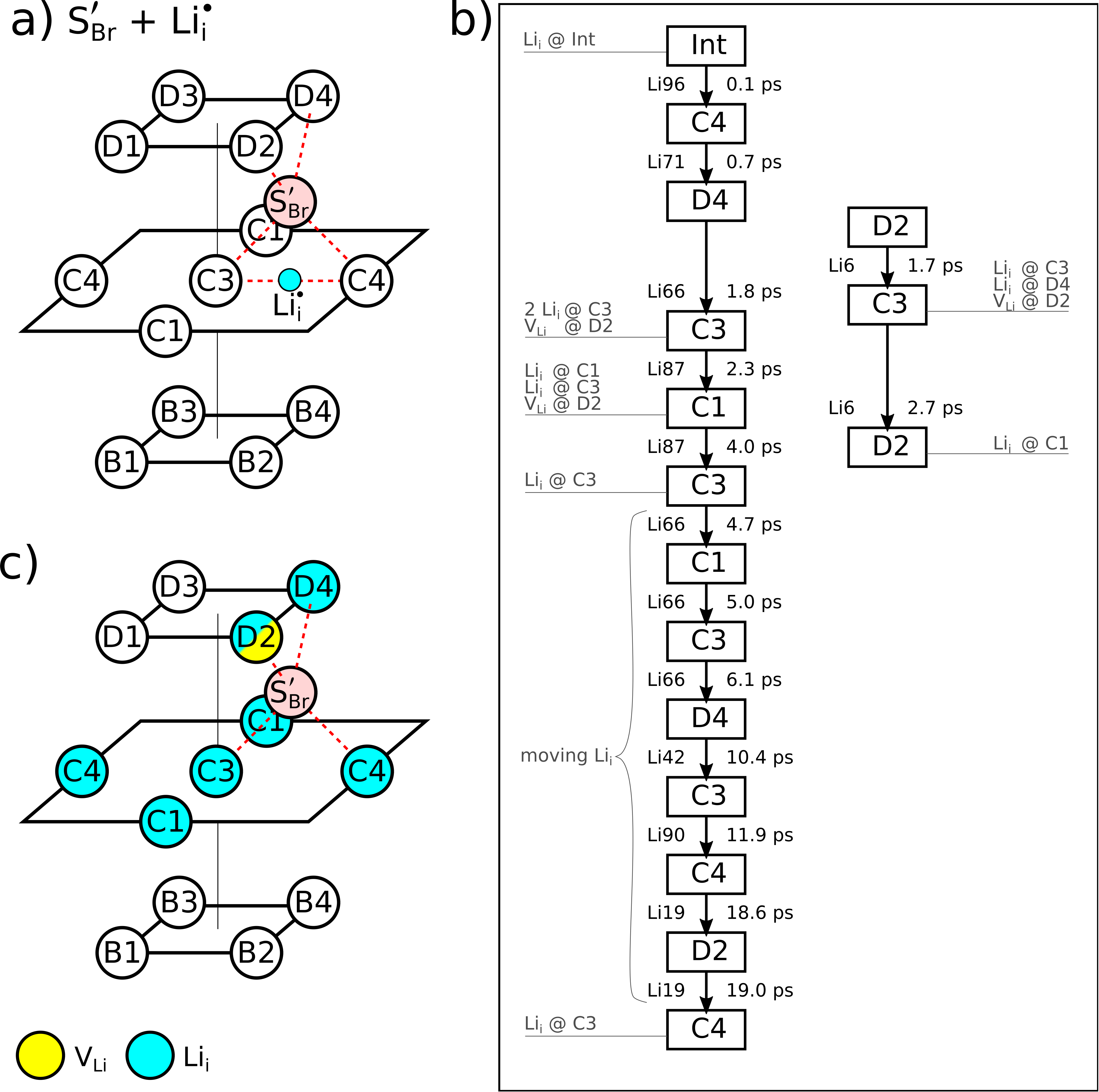}
    \caption{a) Sketch of the initial configuration for the \ch{S_{Br}'} defect (red sphere). The additionally added Li used for charge compensation is indicated with a small sphere in light blue. b) Flow chart of the observed intercage jumps between the different Li cages for a simulation of approximately 20~ps at 600~K. Next to most of the stages the current situation, i.e, the number and position of Li interstitials and vacancies, is summarized. c) Summarizing sketch indicating around which $4d$ sites interstitial Li and/or Li vacancy have been observed during the simulation. Note that the charge superscript for the Kr\"oger-Vink notation was omitted.}
    \label{fig:S-defect}
\end{figure}

A \ch{S_{Br}'} defect was positioned in the supercell structure on the $4a$ site which is tetrahedrally coordinated by the C3, C4, D2 and D4 sites as shown in Figure~\ref{fig:S-defect}a.
The charge compensation was ensured by adding a Li to an interstitial site (\ch{Li_i^.}) in the vicinity of the defect between C3 and C4.

The flow chart shown in Figure~\ref{fig:S-defect}b depicts all observed intercage jumps during the simulation.
Immediately after the simulation was started the \ch{Li_i^.} moves towards C4 and only 0.6~ps later another Li$^+$ jumps from C4 to D4 which is similar to an interstitialcy mechanism.
Furthermore, an additional Frenkel pair defect is formed by a Li$^+$ jump from D2 to C3 but is annihilated again only 1~ps afterwards.
In the meantime, another Li$^+$ jumps from D4 to C3 creating a situation in which even two \ch{Li_i^.} are residing around C3 for a short time until one Li$^+$ moves to the next cage and D2 is refilled. 
For the rest of the simulation only one Li$^+$ is traveling at a time via interstital and interstitialcy jumps.

In conclusion, the majority of Li$^+$ intercage jumps can be attributed to interstitials that jump between the cages.
Because most jumps involve the cages C3, C4, D2 and D4 we find that the interstitial mainly resides in the vicinity of the \ch{S_{Br}'} defect which is summarized in Figure~\ref{fig:S-defect}c.
Both observations are not unexpected: compared to ordinary $4a$ sites occupied with Br$^-$ the \ch{S_{Br}'} defect can be viewed as a negative charge that will more strongly attract the positively charged Li$^+$. 
Therefore, the interstitial, which can be viewed as an additional positive charge, will preferentially stay in the orbit of \ch{S_{Br}'}.
However, also second nearest neighbor cages (here C1) have been reached by the interstitial sporadically.

\subsection{Li diffusion induced by Br on $4d$}

\begin{figure}[ht!]
    \centering
    \includegraphics[width=0.75\textwidth]{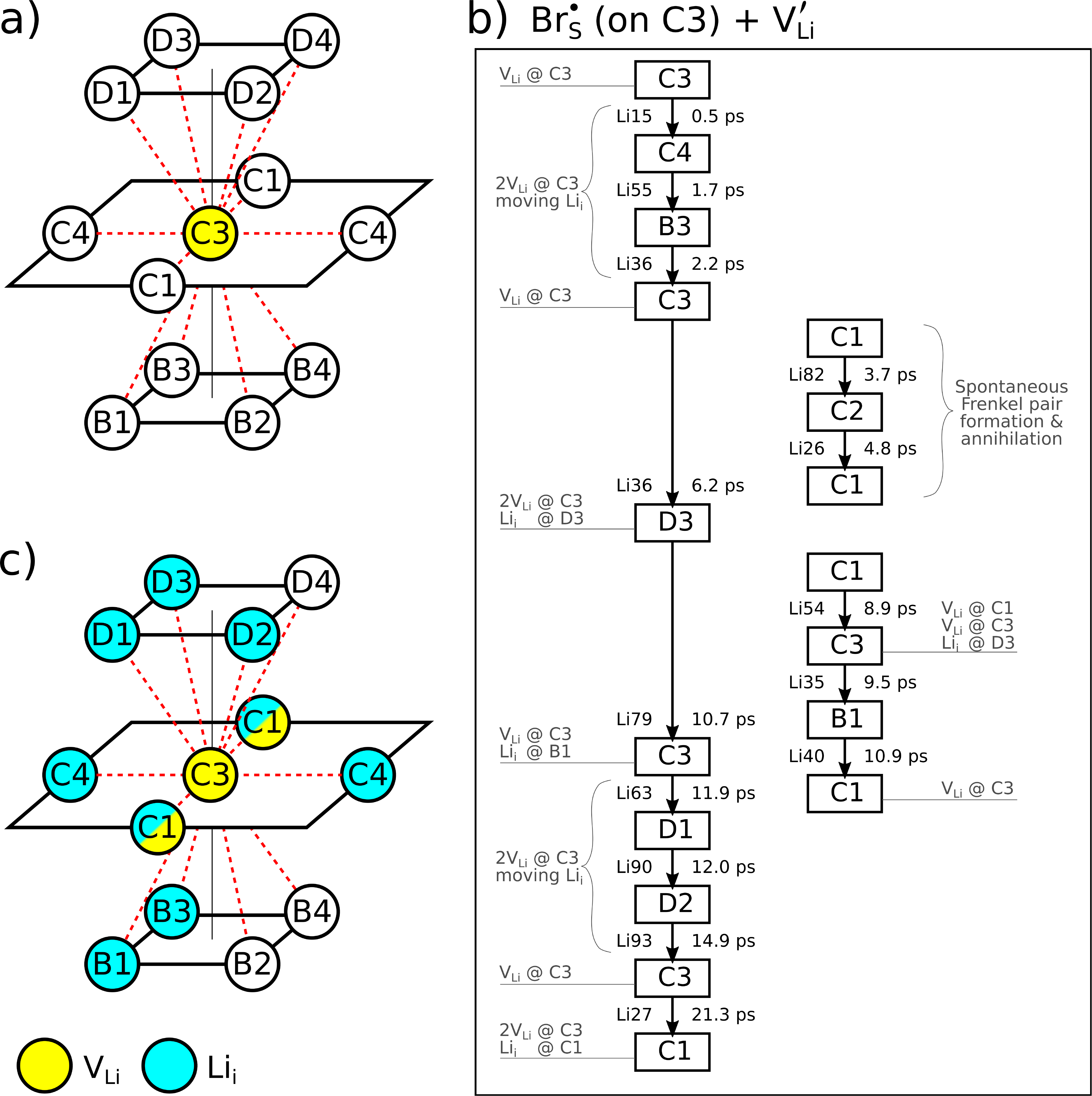}
    \caption{a) Sketch of the initial setup where the $4d$ site on the C3 position is occupied with an \ch{Br_S^.} defect and one Li ion around C3 has been removed for charge compensation. b) Flow chart of the observed intercage jumps between the different Li cages for a simulation of approximately 20~ps at 600~K. Next to most of the stages the current situation, i.e, the number and position of Li interstitials and vacancies, is summarized. c) Summarizing sketch indicating around which $4d$ sites interstitial Li and Li vacancy have been observed during the simulation. For several periods of time even two vacancies are found around C3.}
    \label{fig:Br-defect}
\end{figure}

Without changing the composition of Li$_6$PS$_5$Br the natural counterpart of a \ch{S_{Br}'} defect is \ch{Br_S^.}.
In order to study its effect on the transport properties separately one $4d$ site (C3) was occupied with Br instead of the nominal S.
A Li vacancy \ch{V_{Li}'} was created by removing one Li from the cage around C3 to ensure charge neutrality for the starting structure.
This initial configuration is depicted in Figure~\ref{fig:Br-defect}a together with the corresponding flow chart of the observed intercage jump events as shown in Figure~\ref{fig:Br-defect}b.

At the beginning of the simulation the first event is performed by a Li$^+$ jumping from C3 to C4 thereby creating an interstitial around C4 and leaving back a second vacancy on C3.
Two subsequent events later the interstitial has returned to C3 via interstitialcy jumps.
Overall, the majority of intercage jumps are governed by moving interstitials and during the simulation of approximately 20~ps Li$^+$ interstitials were found in the cages around B1, B3, C1, C4, D1, D2 and D3 as summarized in Figure~\ref{fig:Br-defect}c.
These cages are the nearest neighbor cages around C3 that is hosting the \ch{Br_S^.} defect.
In another independent simulation with the identical initial setup also the remaining nearest neighbor cages B2, B4 and D4 have been visited by an interstitial (see Figure~S1 in the Supplementary Material).

During the simulation shown in Figure~\ref{fig:Br-defect} only in one other cage (C2), not directly next to C3, an interstitial was found. 
This interstitial was generated as part of a spontaneous Frenkel pair together with a vacancy on C1.
Similar to the Frenkel pair observed above for the \ch{S_{Br}'} defect the lifetime again is approximately 1~ps after which it was annihilated again.
Later at around 8.9~ps, when again two vacancies have been present on C3, a Li$^+$ jumped from C1 to C3 to refill one of these vacancies.
This is the only jump event which, instead of a interstitial jump mechanism, could likewise be regarded as a vacancy jump mechanism.

In summary, we found that a \ch{Br_S^.} defect facilitates the generation of Frenkel pairs with rather mobile lithium interstitials and vacancies which are strongly bound to the Li cage around the \ch{Br_S^.} defect. 
The Frenkel pair is formed regardless of the fact that a Li vacancy has already been introduced to the system as compensating defect for \ch{Br_S^.}.
The reason for this is most probably due to the electrostatics: the Br$^-$ on the $4d$ site has a weaker attraction for Li$^+$ compared to the nominal S$^{2-}$.
Moreover, the strong attraction of S$^{2-}$ on the surrounding $4d$ sites is even able to strip a Li$^+$ from the Li cage around \ch{Br_S^.} leaving back two vacancies.
The generated interstitial is highly mobile which is most likely due to a combination of high Li mobility inside the Li cages and structural frustration~\cite{morgan_chem_2020}.
As a result, the interstitial is able to travel easily through all 12 nearest neighbor cages around \ch{Br_S^.}.
With that an extended spatial area is involved with Li transport only because of one single \ch{Br_S^.} defect.

\subsection{Combining S on $4a$ and Br on $4d$}

The previous subsections revealed the effects of the individual \ch{S_{Br}'} and \ch{Br_S^.} defects on $4a$ and $4d$. 
For the analysis, however, the Li content needed to be adjusted to keep the cell charge neutral which might affect the Li transport.
Therefore, the same analysis as presented above has been applied to supercells containing a defect pair of \ch{S_{Br}'} and \ch{Br_S^.} at the same time which corresponds to a site-exchange of 6.25\% and does not require any further charge compensation.
Within the used supercell only four symmetrically dissimilar configurations can be realized.
These have been chosen in the way that the \ch{S_{Br}'} defect always resides on the tetrahedral site between C3, C4, D1 and D3 while the \ch{Br_S^.} defect occupies A2, A3, C2 or C3.
The corresponding flow charts of 50-55~ps simulations at 600~K are shown in Figure~S2-S5 in the Supplementary Material.
Overall, the observations made for the individual \ch{S_{Br}'} and \ch{Br_S^.} defects still hold true and can be summarized as follows.

At almost any time at least one Li Frenkel pair is present in the supercell.
Similar as observed for the separated \ch{S_{Br}'} defect the formation of the Frenkel pair again mainly originates from the \ch{Br_S^.} defect which donates one Li$^+$ from its cage to a neighboring cage.
While the generated interstitial is highly mobile and initiates further lithium intercage jumps via interstitial and interstitialcy mechanisms the \ch{V'_{Li}} vacancy is strongly bound to the \ch{Br_S^.} defect. 
Periodically, even two Frenkel pairs with two mobile interstitials and two immobile vacancies are found around the \ch{Br_S^.} defect.

The \ch{S_{Br}'} defect can be seen as a sink for the moving interstitial.
Frequently, the mobile interstitial is drawn close to the \ch{S_{Br}'} defect and occupies sites that do not belong to the original cages around the $4d$ sites.
This is the onset of the cage shifting towards the $4a$ sites which becomes more pronounced at higher degrees of site-exchange~\cite{morgan_chem_2020}.
Occasionally, the \ch{S_{Br}'} defect is also able to strip a Li from its neighboring cages but the main source for mobile \ch{Li^._i} interstitials still is the \ch{Br_S^.} defect.

\begin{figure}[ht!]
    \centering
    \includegraphics[width=0.8\textwidth]{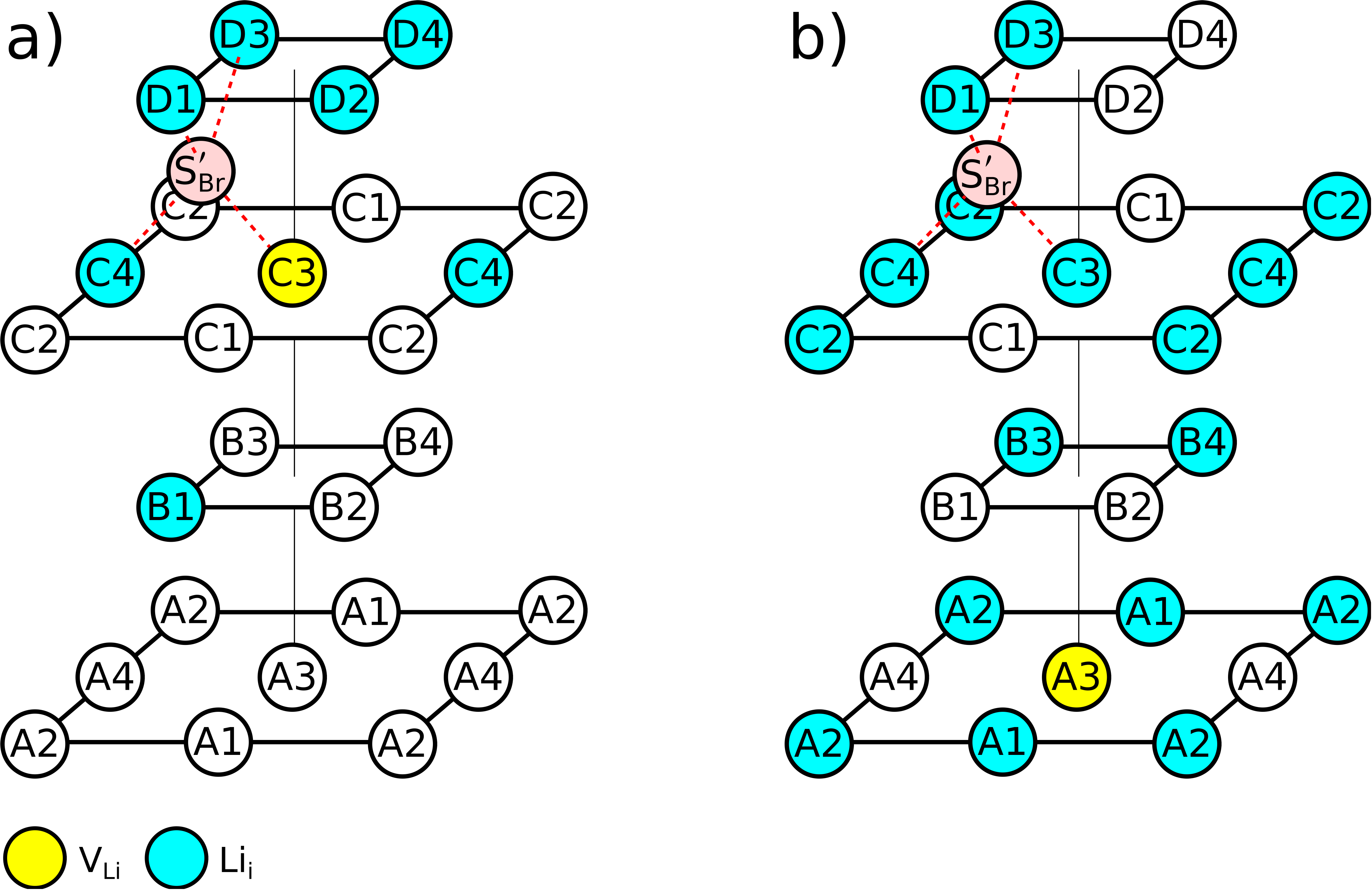}
    \caption{Summarizing sketches indicating around which $4d$ sites interstitial Li and Li vacancies have been observed during the simulation. The position of the \ch{S_{Br}'} defect is the same in both systems and only the \ch{Br_S^.} position varies. For a) it is located on the C3 position whereas for b) the A3 position has been used. The initial structure started with a uniform Li distribution with six Li$^+$ around each $4d$ site.}
    \label{fig:S-Br-distance}
\end{figure}

If the \ch{Br_S^.} and \ch{S_{Br}'} defects are close to each other, the Li intercage jumps are mainly confined to their vicinity as can be seen in Figure~\ref{fig:S-Br-distance}a.
In this setup the two defects are direct neighbors.
As a result, the moving lithium jumps mostly between the cages C4, D1 and D3 or back to C3 where the \ch{Br_S^.} defect is located.
In rare cases the \ch{Li^._i} interstitial also shortly passes the cages B1, D2 and D4 (see Figure~S2 in Supplementary Material).
With that an interstitial was found around 6 of the 15 different $4d$ sites (neglecting C3 which is the 16th site containing the \ch{Br_S^.} defect) during the simulation. 

A larger separation between the two \ch{Br_S^.} and \ch{S_{Br}'} defects, as shown in Figure~\ref{fig:S-Br-distance}b, seems to facilitate a more extensive distribution of the \ch{Li^._i} interstitial.
For this setup interstitials were found around 10 of the 15 different $4d$ sites. 
This observation indicates that in case of low site-exchange the local arrangement of Br$^-$/S$^{2-}$ might play a role for the mobility of Li and its ability to cover long ranges.

\section{Conclusion}
In this work the lithium diffusion processes in the agyrodite-type solid electrolyte Li$_6$PS$_5$Br have been thoroughly analyzed using ab-initio molecular dynamics simulations.
Without Br$^-$/S$^{2-}$ site-exchange lithium diffusion is restricted to ionic motion within the Li$^+$ cages around the $4d$ sites and a new cooperative mechanism for the internal reorganization of lithium was observed.
This reorganization mechanism is highly concerted and cannot be described by one single rotation axis.
Even at elevated temperature of 600~K only one intercage jump was observed during the simulations.
This jump, however, was followed by a back-jump indicating a high correlation.

With the introduction of \ch{Br_S^.} and \ch{S_{Br}'} defects on the $4d$ and $4a$ sites much more intercage jumps are initiated.
Both defects have been analyzed separately and in combination.
From these simulations we identified \ch{Li^._i} interstitials as the main mobile charge carriers.
They originate from Frenkel pairs which are formed mainly because of \ch{Br_S^.} defects on the $4d$ sites that donate one or even two \ch{Li^._i} interstitials to the neighboring 12 cages.
The \ch{Li^._i} interstitials then carry out intercage jumps via interstitial and interstitialcy mechanisms.
With that, one single \ch{Br_S^.} defect enables Li diffusion over an extended spatial area explaining why low degrees of site-exchange are sufficient to trigger superionic conduction. 
The \ch{V_{Li}'} vacancies that are formed as part of the Frenkel pairs are mostly immobile and bound to the \ch{Br_S^.} defect.
To a lesser degree also \ch{S_{Br}'} defects seem to induce disturbances in the lithium distribution and act as sinks for  \ch{Li^._i} interstitials which mostly restricts the lithium motion to the vicinity of the \ch{S_{Br}'} defect. 

\section*{Acknowledgement}
The authors appreciate funding from \textit{Bundesministerium f\"ur Bildung und For\-schung} (BMBF) under grant 03XP0174A.
Calculations for this research were conducted on the Lichtenberg high performance computer of the Technical University of Darmstadt.

\bibliographystyle{rsta}
\bibliography{mybibfile}

\begin{thebibliography}{10}
\expandafter\ifx\csname urlstyle\endcsname\relax
  \providecommand{\doi}[1]{(doi:\discretionary{}{}{}#1)}\else
  \providecommand{\doi}{(doi:\discretionary{}{}{}\begingroup
  \urlstyle{rm}\Url)}\fi

\bibitem{deiseroth_li6ps5x_2008}
Deiseroth HJ, Kong ST, Eckert H, Vannahme J, Reiner C, Zai\ss\ T, Schlosser M.
  2008 {Li6PS5X}: {A} {Class} of {Crystalline} {Li}-{Rich} {Solids} {With} an
  {Unusually} {High} {Li}+ {Mobility}.
\newblock \emph{Angewandte Chemie International Edition} \textbf{47}, 755--758.
\newblock \doi{10.1002/anie.200703900}

\bibitem{deiseroth_li7ps6_2011}
Deiseroth HJ, Maier J, Weichert K, Nickel V, Kong ST, Reiner C. 2011 {Li7PS6}
  and {Li6PS5X} ({X}: {Cl}, {Br}, {I}): {Possible} {Three}-dimensional
  {Diffusion} {Pathways} for {Lithium} {Ions} and {Temperature} {Dependence} of
  the {Ionic} {Conductivity} by {Impedance} {Measurements}.
\newblock \emph{Zeitschrift f\"ur anorganische und allgemeine Chemie}
  \textbf{637}, 1287--1294.
\newblock \doi{10.1002/zaac.201100158}

\bibitem{kraft_influence_2017}
Kraft MA, Culver SP, Calderon M, B\"ocher F, Krauskopf T, Senyshyn A, Dietrich
  C, Zevalkink A, Janek J, Zeier WG. 2017 Influence of {Lattice}
  {Polarizability} on the {Ionic} {Conductivity} in the {Lithium} {Superionic}
  {Argyrodites} {Li6PS5X} ({X} = {Cl}, {Br}, {I}).
\newblock \emph{Journal of the American Chemical Society} \textbf{139},
  10909--10918.
\newblock \doi{10.1021/jacs.7b06327}

\bibitem{yu_accessing_2017}
Yu C, Ganapathy S, Eck ERHv, Wang H, Basak S, Li Z, Wagemaker M. 2017 Accessing
  the bottleneck in all-solid state batteries, lithium-ion transport over the
  solid-electrolyte-electrode interface.
\newblock \emph{Nature Communications} \textbf{8}, 1--9.
\newblock \doi{10.1038/s41467-017-01187-y}

\bibitem{hanghofer_fast_2019}
Hanghofer I, Gadermaier B, Wilkening HMR. 2019 Fast {Rotational} {Dynamics} in
  {Argyrodite}-{Type} {Li6PS5X} ({X}: {Cl}, {Br}, {I}) as {Seen} by {31P}
  {Nuclear} {Magnetic} {Relaxation} -- {On} {Cation}-{Anion} {Coupled}
  {Transport} in {Thiophosphates}.
\newblock \emph{Chemistry of Materials} \textbf{31}, 4591--4597.
\newblock \doi{10.1021/acs.chemmater.9b01435}

\bibitem{minafra_local_2020}
Minafra N, Kraft MA, Bernges T, Li C, Schlem R, Morgan BJ, Zeier WG. 2020 Local
  {Charge} {Inhomogeneity} and {Lithium} {Distribution} in the {Superionic}
  {Argyrodites} {Li6PS5X} ({X} = {Cl}, {Br}, {I}).
\newblock \emph{Inorganic Chemistry} \textbf{59}, 11009--11019.
\newblock \doi{10.1021/acs.inorgchem.0c01504}.
\newblock Publisher: American Chemical Society

\bibitem{morgan_chem_2020}
Morgan B. 2020 {Mechanistic} {Origin} of {Superionic} {Lithium} {Diffusion} in
  {Anion}-{Disordered} {Li6PS5X} {Argyrodites}.
\newblock \emph{Chemistry of Materials} .\doi{10.26434/chemrxiv.12349703.v1}

\bibitem{yu_unravelling_2016}
Yu C, Ganapathy S, de~Klerk NJJ, Roslon I, van Eck ERH, Kentgens APM, Wagemaker
  M. 2016 Unravelling {Li}-{Ion} {Transport} from {Picoseconds} to {Seconds}:
  {Bulk} versus {Interfaces} in an {Argyrodite} {Li6PS5Cl}-{Li2S}
  {All}-{Solid}-{State} {Li}-{Ion} {Battery}.
\newblock \emph{Journal of the American Chemical Society} \textbf{138},
  11192--11201.
\newblock \doi{10.1021/jacs.6b05066}

\bibitem{chen_stability_2015}
Chen HM, Maohua C, Adams S. 2015 Stability and ionic mobility in
  argyrodite-related lithium-ion solid electrolytes.
\newblock \emph{Physical Chemistry Chemical Physics} \textbf{17}, 16494--16506.
\newblock \doi{10.1039/C5CP01841B}

\bibitem{wang_theoretical_2017}
Wang Z, Shao G. 2017 Theoretical design of solid electrolytes with superb ionic
  conductivity: alloying effect on {Li}+ transportation in cubic {Li6PA5X}
  chalcogenides.
\newblock \emph{Journal of Materials Chemistry A} \textbf{5}, 21846--21857.
\newblock \doi{10.1039/C7TA06986C}

\bibitem{zhu_li3yps42_2017}
Zhu Z, Chu IH, Ong SP. 2017 {Li3Y}({PS4})2 and {Li5PS4Cl2}: {New} {Lithium}
  {Superionic} {Conductors} {Predicted} from {Silver} {Thiophosphates} using
  {Efficiently} {Tiered} {Ab} {Initio} {Molecular} {Dynamics} {Simulations}.
\newblock \emph{Chemistry of Materials} \textbf{29}, 2474--2484.
\newblock \doi{10.1021/acs.chemmater.6b04049}

\bibitem{adeli_boosting_2019}
Adeli P, Bazak JD, Park KH, Kochetkov I, Huq A, Goward GR, Nazar LF. 2019
  Boosting {Solid}-{State} {Diffusivity} and {Conductivity} in {Lithium}
  {Superionic} {Argyrodites} by {Halide} {Substitution}.
\newblock \emph{Angewandte Chemie International Edition} \textbf{58},
  8681--8686.
\newblock \doi{10.1002/anie.201814222}

\bibitem{ouyang_computational_2020}
Ouyang B, Wang Y, Sun Y, Ceder G. 2020 Computational {Investigation} of
  {Halogen}-{Substituted} {Na} {Argyrodites} as {Solid}-{State} {Superionic}
  {Conductors}.
\newblock \emph{Chemistry of Materials} \textbf{32}, 1896--1903.
\newblock \doi{10.1021/acs.chemmater.9b04541}.
\newblock Publisher: American Chemical Society

\bibitem{wang_fast_2020}
Wang P, Liu H, Patel S, Feng X, Chien PH, Wang Y, Hu YY. 2020 Fast {Ion}
  {Conduction} and {Its} {Origin} in {Li6}-{xPS5}-{xBr1}+x.
\newblock \emph{Chemistry of Materials} \textbf{32}, 3833--3840.
\newblock \doi{10.1021/acs.chemmater.9b05331}.
\newblock Publisher: American Chemical Society

\bibitem{zhang_silicon-doped_2020}
Zhang J, Li L, Zheng C, Xia Y, Gan Y, Huang H, Liang C, He X, Tao X, Zhang W.
  2020 Silicon-{Doped} {Argyrodite} {Solid} {Electrolyte} {Li6PS5I} with
  {Improved} {Ionic} {Conductivity} and {Interfacial} {Compatibility} for
  {High}-{Performance} {All}-{Solid}-{State} {Lithium} {Batteries}.
\newblock \emph{ACS Applied Materials \& Interfaces}
  .\doi{10.1021/acsami.0c11683}.
\newblock Publisher: American Chemical Society

\bibitem{gautam_rapid_2019}
Gautam A, Sadowski M, Prinz N, Eickhoff H, Minafra N, Ghidiu M, Culver SP, Albe
  K, F{\"a}ssler TF, Zobel M, Zeier WG. 2019 Rapid {Crystallization} and
  {Kinetic} {Freezing} of {Site}-{Disorder} in the {Lithium} {Superionic}
  {Argyrodite} {Li6PS5Br}.
\newblock \emph{Chemistry of Materials} \textbf{31}, 10178--10185.
\newblock \doi{10.1021/acs.chemmater.9b03852}

\bibitem{pecher_atomistic_2010}
Pecher O, Kong ST, Goebel T, Nickel V, Weichert K, Reiner C, Deiseroth HJ,
  Maier J, Haarmann F, Zahn D. 2010 Atomistic {Characterisation} of {Li}+
  {Mobility} and {Conductivity} in {Li7}-{xPS6}-{xIx} {Argyrodites} from
  {Molecular} {Dynamics} {Simulations}, {Solid}-{State} {NMR}, and {Impedance}
  {Spectroscopy}.
\newblock \emph{Chemistry -- A European Journal} \textbf{16}, 8347--8354.
\newblock \doi{10.1002/chem.201000501}

\bibitem{rao_studies_2011}
Rao RP, Adams S. 2011 Studies of lithium argyrodite solid electrolytes for
  all-solid-state batteries.
\newblock \emph{physica status solidi (a)} \textbf{208}, 1804--1807.
\newblock \doi{10.1002/pssa.201001117}

\bibitem{rayavarapu_variation_2012}
Rayavarapu PR, Sharma N, Peterson VK, Adams S. 2012 Variation in structure and
  {Li}+-ion migration in argyrodite-type {Li6PS5X} ({X} = {Cl}, {Br}, {I})
  solid electrolytes.
\newblock \emph{Journal of Solid State Electrochemistry} \textbf{16},
  1807--1813.
\newblock \doi{10.1007/s10008-011-1572-8}

\bibitem{holzwarth_computer_2011}
Holzwarth NAW, Lepley ND, Du YA. 2011 Computer modeling of lithium phosphate
  and thiophosphate electrolyte materials.
\newblock \emph{Journal of Power Sources} \textbf{196}, 6870--6876.
\newblock \doi{10.1016/j.jpowsour.2010.08.042}

\bibitem{de_klerk_diffusion_2016}
de~Klerk NJJ, Ros{\l}o\'n I, Wagemaker M. 2016 Diffusion {Mechanism} of {Li}
  {Argyrodite} {Solid} {Electrolytes} for {Li}-{Ion} {Batteries} and
  {Prediction} of {Optimized} {Halogen} {Doping}: {The} {Effect} of {Li}
  {Vacancies}, {Halogens}, and {Halogen} {Disorder}.
\newblock \emph{Chemistry of Materials} \textbf{28}, 7955--7963.
\newblock \doi{10.1021/acs.chemmater.6b03630}

\bibitem{deng_data-driven_2017}
Deng Z, Zhu Z, Chu IH, Ong SP. 2017 Data-{Driven} {First}-{Principles}
  {Methods} for the {Study} and {Design} of {Alkali} {Superionic} {Conductors}.
\newblock \emph{Chemistry of Materials} \textbf{29}, 281--288.
\newblock \doi{10.1021/acs.chemmater.6b02648}

\bibitem{stamminger_ionic_2019}
Stamminger AR, Ziebarth B, Mrovec M, Hammerschmidt T, Drautz R. 2019 Ionic
  {Conductivity} and {Its} {Dependence} on {Structural} {Disorder} in
  {Halogenated} {Argyrodites} {Li6PS5X} ({X} = {Br}, {Cl}, {I}).
\newblock \emph{Chemistry of Materials} \textbf{31}, 8673--8678.
\newblock \doi{10.1021/acs.chemmater.9b02047}

\bibitem{vasp1}
Kresse G, Hafner J. 1994 Ab initio molecular-dynamics simulation of the
  liquid-metal-amorphous-semiconductor transition in germanium.
\newblock \emph{Physical Review B} \textbf{49}, 14251--14269.
\newblock \doi{10.1103/PhysRevB.49.14251}

\bibitem{vasp2}
Kresse G. 1995 Ab initio molecular dynamics for liquid metals.
\newblock \emph{Journal of Non-Crystalline Solids} \textbf{192-193}, 222--229.
\newblock \doi{10.1016/0022-3093(95)00355-X}

\bibitem{vasp3}
Kresse G, Furthm{\"u}ller J. 1996 Efficiency of ab-initio total energy
  calculations for metals and semiconductors using a plane-wave basis set.
\newblock \emph{Computational Materials Science} \textbf{6}, 15--50.
\newblock \doi{10.1016/0927-0256(96)00008-0}

\bibitem{vasp4}
Kresse G, Furthm{\"u}ller J. 1996 Efficient iterative schemes for ab initio
  total-energy calculations using a plane-wave basis set.
\newblock \emph{Physical Review B} \textbf{54}, 11169--11186.
\newblock \doi{10.1103/PhysRevB.54.11169}

\bibitem{pbe1}
Perdew JP, Burke K, Ernzerhof M. 1996 Generalized {Gradient} {Approximation}
  {Made} {Simple}.
\newblock \emph{Physical Review Letters} \textbf{77}, 3865--3868.
\newblock \doi{10.1103/PhysRevLett.77.3865}

\bibitem{pbe2}
Perdew JP, Burke K, Ernzerhof M. 1997 Generalized {Gradient} {Approximation}
  {Made} {Simple} [{Phys}. {Rev}. {Lett}. 77, 3865 (1996)].
\newblock \emph{Physical Review Letters} \textbf{78}, 1396--1396.
\newblock \doi{10.1103/PhysRevLett.78.1396}

\bibitem{paw1}
Bl{\"o}chl PE. 1994 Projector augmented-wave method.
\newblock \emph{Physical Review B} \textbf{50}, 17953--17979.
\newblock \doi{10.1103/PhysRevB.50.17953}

\bibitem{paw2}
Kresse G, Joubert D. 1999 From ultrasoft pseudopotentials to the projector
  augmented-wave method.
\newblock \emph{Physical Review B} \textbf{59}, 1758--1775.
\newblock \doi{10.1103/PhysRevB.59.1758}

\bibitem{okhotnikov_supercell_2016}
Okhotnikov K, Charpentier T, Cadars S. 2016 Supercell program: a combinatorial
  structure-generation approach for the local-level modeling of atomic
  substitutions and partial occupancies in crystals.
\newblock \emph{Journal of Cheminformatics} \textbf{8}, 17.
\newblock \doi{10.1186/s13321-016-0129-3}

\bibitem{momma_vesta_2011}
Momma K, Izumi F. 2011 {VESTA} 3 for three-dimensional visualization of
  crystal, volumetric and morphology data.
\newblock \emph{Journal of Applied Crystallography} \textbf{44}, 1272--1276.
\newblock \doi{10.1107/S0021889811038970}

\bibitem{stukowski_visualization_2010}
Stukowski A. 2010 Visualization and analysis of atomistic simulation data with
  {OVITO}-the {Open} {Visualization} {Tool}.
\newblock \emph{Modelling and Simulation in Materials Science and Engineering}
  \textbf{18}, 015012.
\newblock \doi{10.1088/0965-0393/18/1/015012}

\end{thebibliography}


\begin{thebibliography}{}
\expandafter\ifx\csname url\endcsname\relax
  \def\url#1{\texttt{#1}}\fi
\expandafter\ifx\csname urlprefix\endcsname\relax\def\urlprefix{URL }\fi
\expandafter\ifx\csname href\endcsname\relax
  \def\href#1#2{#2} \def\path#1{#1}\fi

\end{thebibliography}

\end{document}


\begin{center}

\Large - Supplementary Material - \\
\LARGE Influence of Br$^{-}$/S$^{2-}$ site-exchange on Li diffusion mechanism in Li$_6$PS$_5$Br - a computational study

\bigskip

\normalsize
Marcel Sadowski and Karsten Albe\\
Email: sadowski@mm.tu-darmstadt.de and albe@mm.tu-darmstadt.de

\bigskip
\
Institute of Materials Science, Technical University of Darmstadt, 64287 Darmstadt, Germany

\end{center}

\bigskip

\bigskip

\begin{figure}[ht!]
    \centering
    \includegraphics[width=0.75\textwidth]{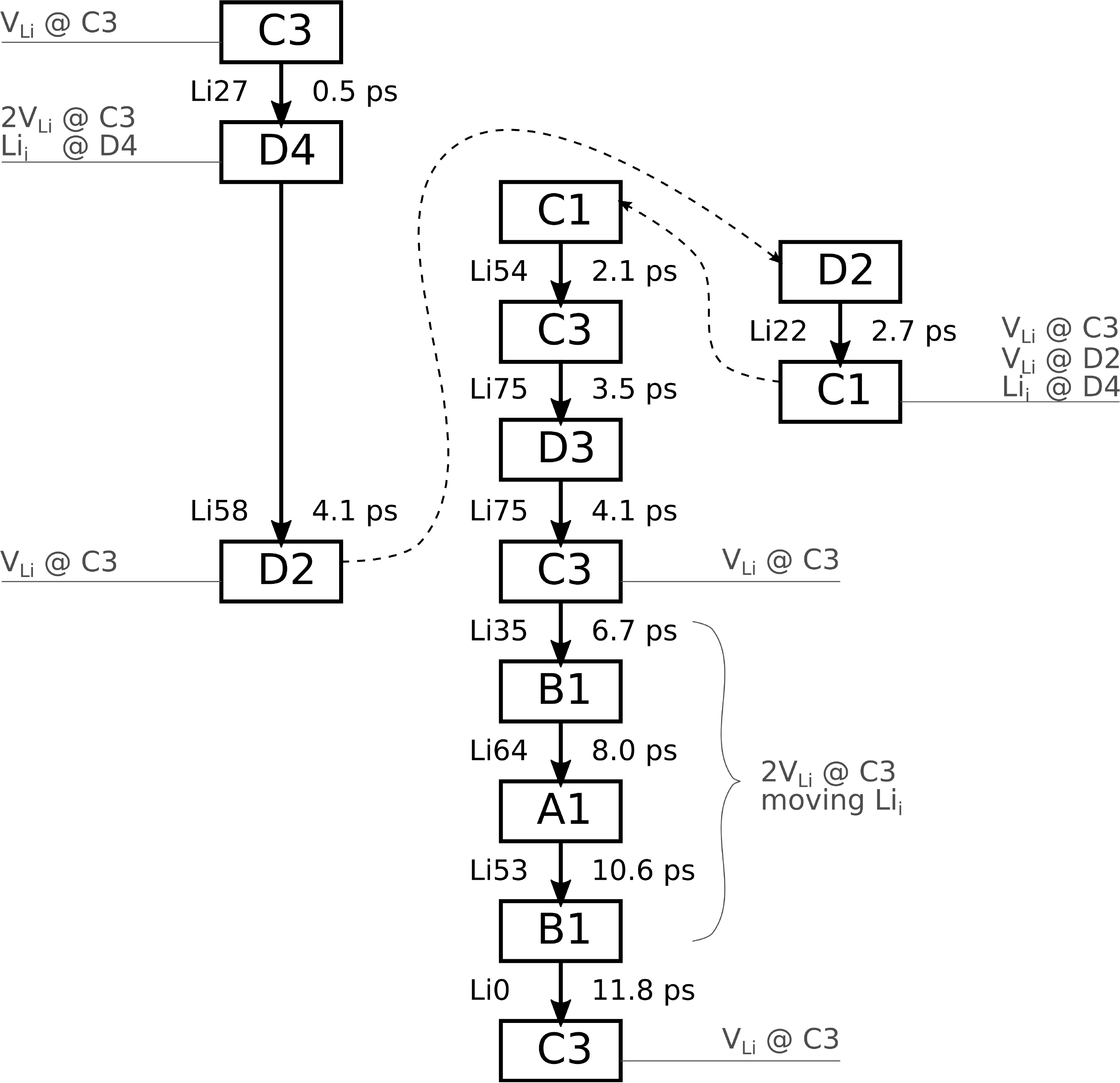}
    \caption{Flow chart of the observed intercage jumps between the different Li cages for a setup with \ch{Br_S^.} defect on the C3 site and an additional Li$^+$ vacancy in its vicinity for the initial structure. Arrows with dashed lines are used to indicate refilling of vacancies.}
    \label{fig:01_flowchart}
\end{figure}

\begin{figure}[ht!]
    \centering
    \includegraphics[width=0.95\textwidth]{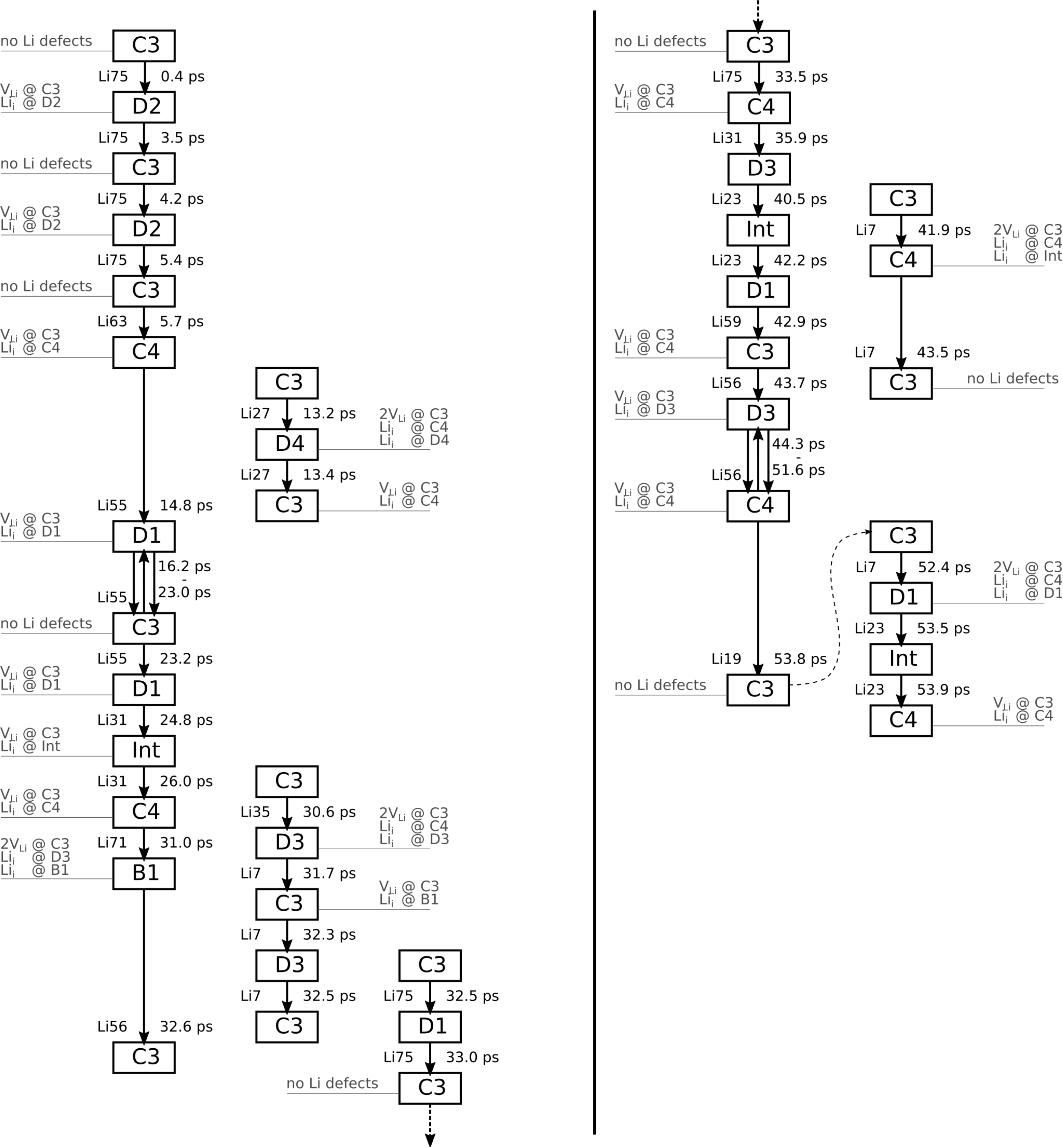}
    \caption{Flow chart of the observed intercage jumps between the different Li cages for a setup with a \ch{S_{Br}'} defect on the tetrahedral site between C3, C4 , D1 and D3, and a \ch{Br_S^.} defect on the C3 site. An arrow with dashed line is used to indicate refilling of a vacancy.}
    \label{fig:01_flowchart}
\end{figure}

\begin{figure}[ht!]
    \centering
    \includegraphics[width=0.95\textwidth]{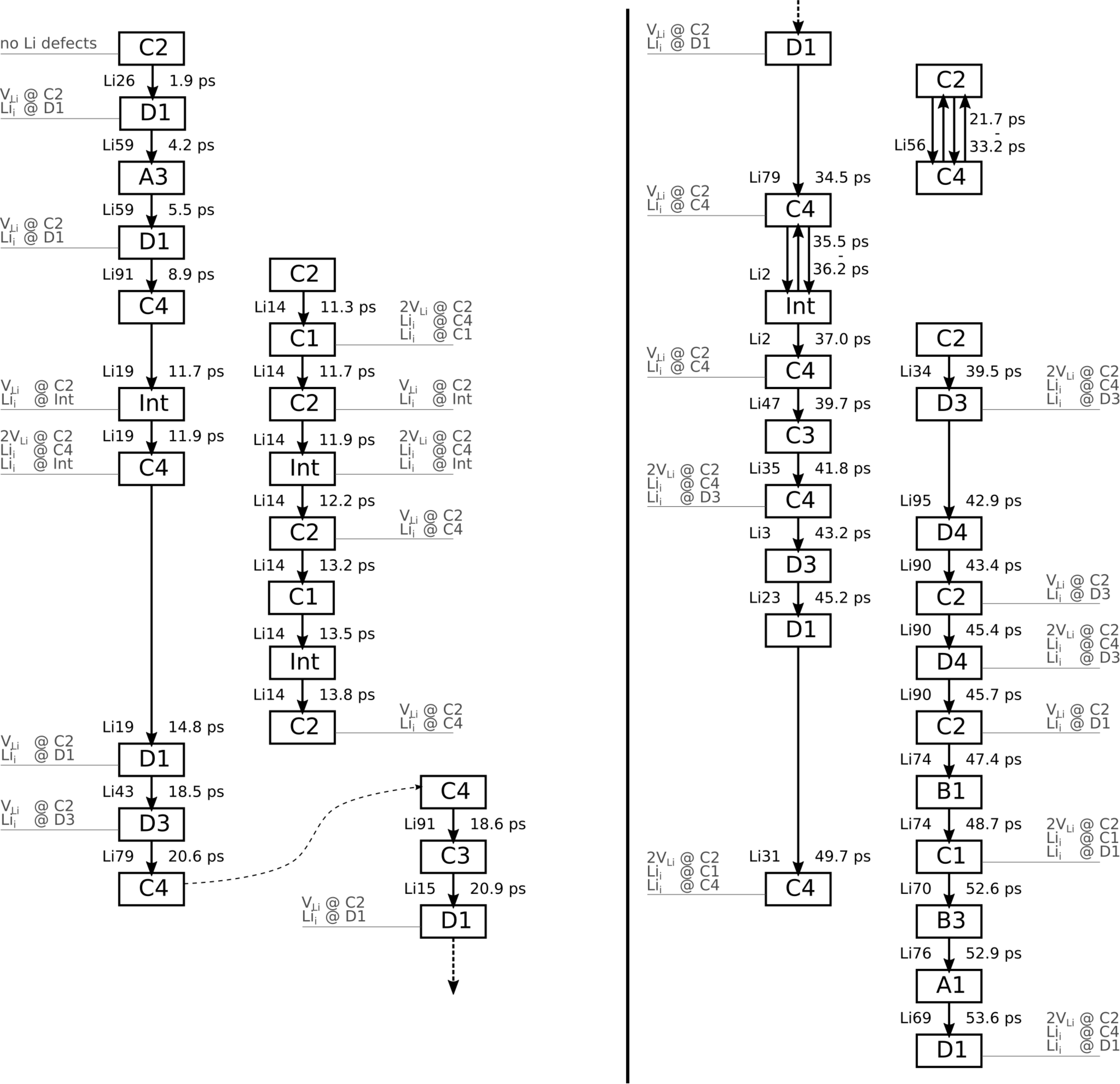}
    \caption{Flow chart of the observed intercage jumps between the different Li cages for a setup with a \ch{S_{Br}'} defect on the tetrahedral site between C3, C4 , D1 and D3, and a \ch{Br_S^.} defect on the C2 site. An arrow with dashed line is used to indicate refilling of a vacancy.}
    \label{fig:02_flowchart}
\end{figure}

\begin{figure}[ht!]
    \centering
    \includegraphics[width=0.95\textwidth]{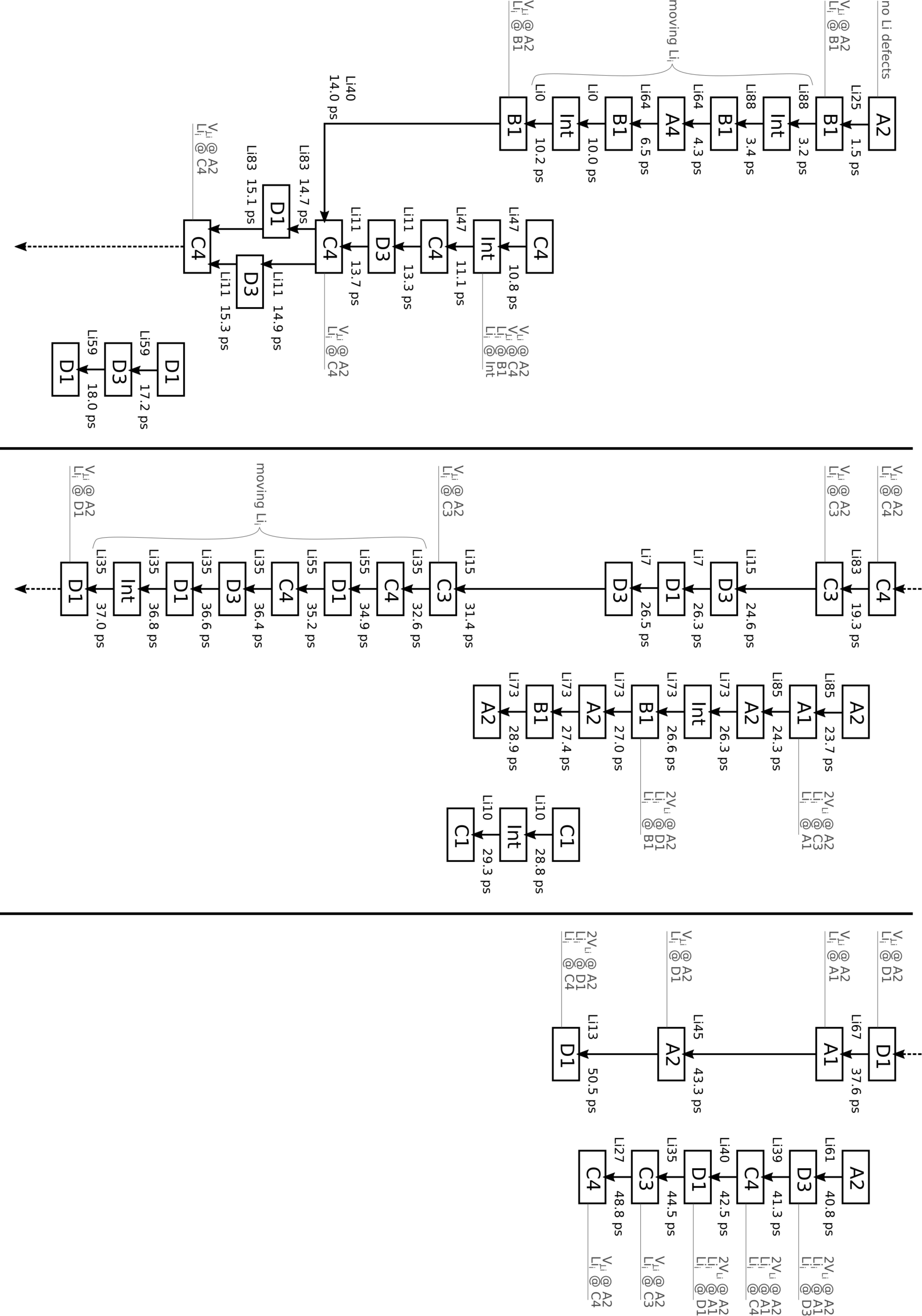}
    \caption{Flow chart of the observed intercage jumps between the different Li cages for a setup with a \ch{S_{Br}'} defect on the tetrahedral site between C3, C4 , D1 and D3, and a \ch{Br_S^.} defect on the A2 site.}
    \label{fig:03_flowchart}
\end{figure}

\begin{figure}[ht!]
    \centering
    \includegraphics[width=0.95\textwidth]{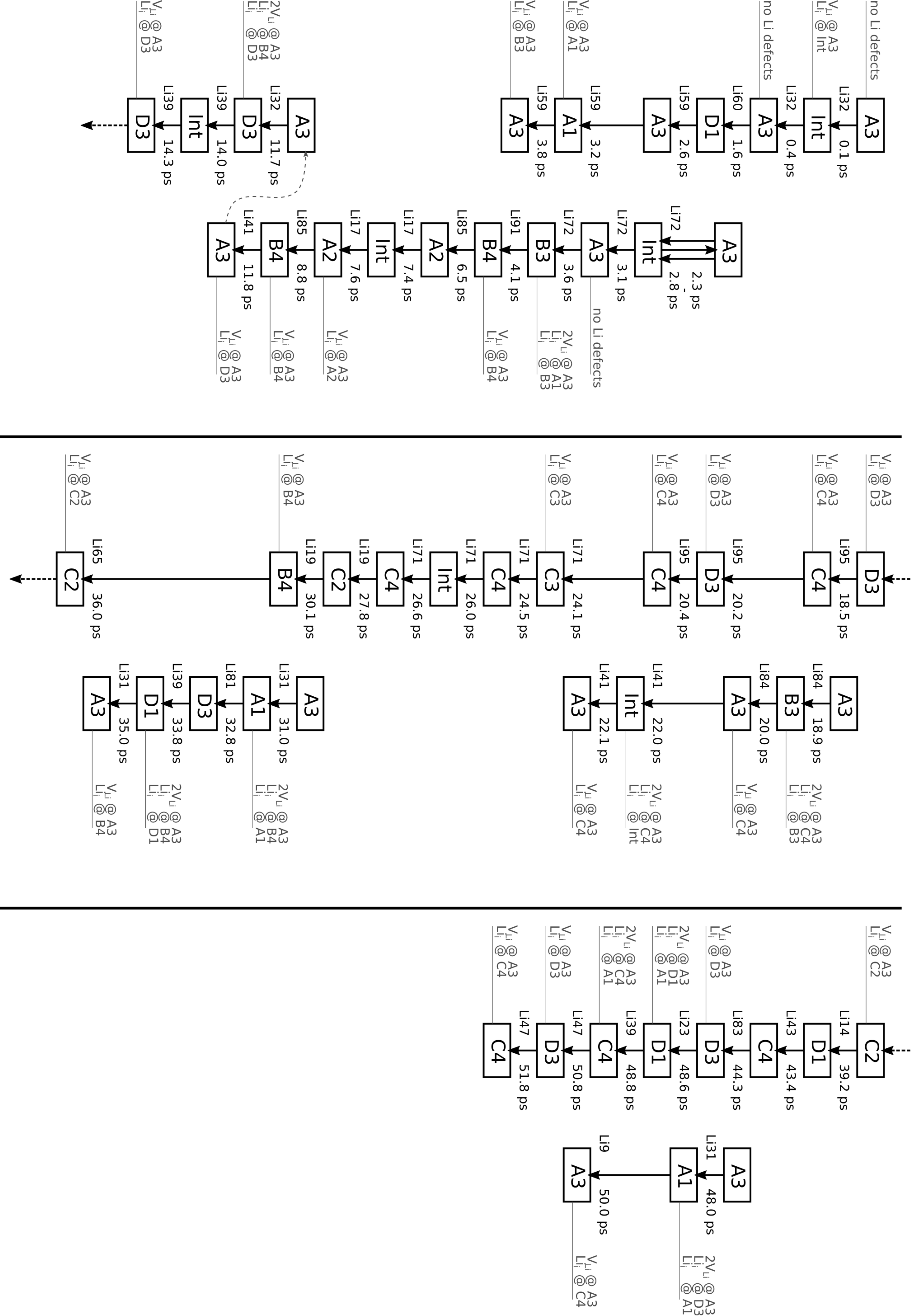}
    \caption{Flow chart of the observed intercage jumps between the different Li cages for a setup with a \ch{S_{Br}'} defect on the tetrahedral site between C3, C4 , D1 and D3, and a \ch{Br_S^.} defect on the A3 site.}
    \label{fig:04_flowchart}
\end{figure}